# Far infrared and terahertz spectroscopy of ferroelectric soft modes in thin films: A review

J. Petzelt and S. Kamba
Institute of Physics, Czech Academy of Sciences
Na Slovance 2, 182 21 Prague 8, Czech Republic

**Abstract**

Far-infrared and terahertz spectroscopy of ferroelectric soft and central modes in thin films on substrates is reviewed. In addition to classical displacive proper ferroelectrics, also incipient and relaxor ferroelectrics and multiferroics are discussed. Special attention is paid to differences between the soft-mode behavior in thin films and bulk materials (ceramics and single crystals) and their influence on the low-frequency permittivity. Particularly the effects of the thin film strains and depolarizing electric fields of the probing waves on the grain boundaries are emphasized. The soft-mode spectroscopy is shown to be a very sensitive tool to reveal the thin film quality.

**Introduction**

Extensive studies of ferroelectric thin films have been carried out since the beginning of 90-ties, see e.g. review articles by Dawber *et al.* [1] and Setter *et al.* [2], because of their attractive properties in many applications and progress in thin-film processing technologies. However, the general review articles are limited mostly to macroscopic properties and do not concern spectroscopic investigations of thin films devoted to their lattice dynamics and soft phonon mode (SM) behavior. It is well known that the existence of SMs represents the basic feature of ferroelectric (FE) phase transitions (PTs) of the displacive type. They appear as heavily damped temperature dependent phonon excitations active in the far-infrared (FIR) and terahertz (THz) range. In proper and pseudo-proper (weak) ferroelectrics they are IR active in both ferroelectric and paraelectric phases so that the FE PT is revealed as a minimum in the SM-frequency vs. temperature dependence and this represents one of the best manifestations of the phase transition. On the other hand, in order-disorder FEs the role of SM is taken over by a critical relaxation, which appears below the phonon frequency range, usually in the GHz-MHz range, and describes the hopping of ions in some of the disordered (strongly anharmonic) sublattices. However, these are limiting cases and most frequently both displacive and order-disorder dynamic features appear simultaneously, at least close to the PT, with only partial softening of the SM, and the critical relaxation at lower frequencies is then called central mode (CM), since it appears as a central peak in inelastic scattering experiments. Several review articles exist on their appearance and characterization in the bulk (i.e. single crystal and ceramic) FEs, incipient FEs and relaxor FEs [3,4,5,6,7,8]. Concerning the FE SM and CM behavior in thin films, only brief reviews have been published up to date [7,9,10,11] or reviews devoted predominantly to Raman data on perovskite films [12]. In this review we are going to summarize briefly the main thin-film data on FIR-THz spectroscopy of FE SMs and CMs in classical FEs, incipient FEs and relaxors, and complement them by more recent FIR-THz data on their temperature, electric-field (*E*-field), magnetic-field and strain dependences and by SM and CM data on new thin-film materials, mostly of the multiferroic type.

**Experiment and data evaluation**

The FIR measurements based on vacuum Fourier transform spectroscopy (usually > 30 cm$^{-1}$) can be carried out in the transmission and/or specular reflection mode (under near-normal incidence), but for time-domain (~5-80 cm$^{-1}$) or monochromatic backward-wave-oscillator (BWO) terahertz (THz) spectroscopy (~3-30 cm$^{-1}$) the transmission mode is preferential. Since it is advantageous to combine both frequency ranges and possibly transmission and reflection techniques on the same sample, usually dielectric substrates are used. Moreover, our experience shows that the metallic substrates (in the reflection mode) with negative permittivity in the FIR range reduce the sensitivity to the film properties so that films with thicknesses below ~1 μm on metallic substrates are usually not visible. Moreover, to calculate the dielectric function of the film, the accurate dielectric function of the substrate is needed and this is a difficult problem for metals due to their high absorption. Only quite recently a first attempt appeared [13] to analyze the SM response of a PZT film (thickness $d$ = 1165 nm, see below) on a substrate with opaque Pt-layer using the BWO-FTIR in the reflection mode. Let us notice that this differs from Raman measurements, where the metallic substrates have the advantage of preventing any Raman signal from the substrate and therefore avoid the subtraction procedure of the substrate spectra from those of the film. This differs also from the FIR ellipsometry with synchrotron source, which is now well available down to THz frequencies [14,15,16,17], but to our knowledge, for SM spectroscopy on thin films it was so far used only rarely, see the discussion later on. Synchrotron sources [18] with FTIR spectrometers, in principle, can well substitute and even overcome the laboratory FTIR and time-domain THz spectrometers (as well as higher-frequency optical spectrometers), but to our knowledge, they were also not used for the SM studies on thin films.

Choice of a suitable substrate depends not only on its good transparency in the THz range, but also on the method of the film processing and assumed microstructure (polycrystalline or epitaxial). For polycrystalline films the mostly used substrates are sapphire, alumina, MgO, quartz, silica, LaAlO$_3$ and insulating Si; the optimum thickness ranges between 0.3 and 0.5 mm to resolve well the transmittance interferences due to multiple passage of the IR radiation through the substrate. For the same reason, the in-plane optical isotropy of the substrate is preferred, since it makes possible to use the unpolarized radiation. For epitaxial films the substrate has to be single crystalline and its lattice parameter should be compatible to that of the film. This was discussed in the review article on strained epitaxial FE films [19,20], where the following used substrates of perovskite and related structures were listed: YAlO$_3$, LaSrAlO$_4$, LaAlO$_3$, LaSrGaO$_4$, NdGaO$_3$, (LaAlO$_3$)$_{0.29}$-(Sr$_{1/2}$Al$_{1/2}$TaO$_3$)$_{0.71}$ (LSAT), LaGaO$_3$, SrTiO$_3$, AScO$_3$ (A = Dy, Tb, Gd, Eu, Sm, Nd, Pr) and KTaO$_3$. For the reflectance SM studies of thin films, FIR non-transparent substrates are preferential, phonons of the substrates should have small dielectric strengths and their frequencies should be preferably temperature independent. For that reason SrTiO$_3$ (STO) substrates are not suitable for FIR reflectance studies of SMs in thin films. Advantage of FIR reflectance studies is that not only SM, but other polar phonon parameters can be evaluated from the spectra.

Our FIR transmission and reflection measurement were performed using FTIR spectrometer Bruker IFS 113v with pyroelectric deuterated triglycine sulfate (DTGS) detector or He-cooled Si bolometer (for cryogenic measurements). The earlier THz and sub-THz measurements were performed using a BWO spectrometer Epsilon [21] and the more recent time-domain THz measurements on a custom-made spectrometer based on a femtosecond Ti:sapphire laser [10]. In both the latter techniques, the transmitted amplitude is measured together with the phase shift due to the sample, which enables to determine the complex dielectric response without any fitting. Since the techniques are electrode-less, the accuracy is determined only by the noise and precision of the thickness measurements and plane-parallelity of both the film and substrate. Since the FTIR reflectance measurements can be influenced by other systematic geometrical errors, we combine the FTIR spectra (if needed normalized to merge properly with the THz data) with the THz dielectric data and fit them together simultaneously using simple phenomenological models, which satisfy the Kramers-Kronig relations between the real and imaginary part of the dielectric response.

FTIR fitting is carried out separately for the bare substrate and the substrate + film using the well-known Fresnel formulae for the normal transmission and/or reflection of the set of plane-parallel samples [22,23]. In case of the THz data, the calculation of the dielectric response is simplified if the bare substrate thickness is precisely the same as the substrate thickness under the film. For measurements of the substrate FTIR spectra, usually the reflectivity measurements are used and for fitting we usually use the set of generalized damped harmonic oscillators in the factorized form of the dielectric function, whereas for fitting the film usually the simpler sum of classical damped harmonic oscillators is preferable since only the transverse optical (TO) modes of the film are revealed in the spectra [5].

**Ferroelectric and incipient ferroelectric perovskites**

The first FIR-THz measurements of FE SMs were carried out on $PbTiO_3$ (PTO) and $PbZr_{1-x}Ti_xO_3$ ($x$ = 0.47 and 0.25, PZT-47 and PZT-25) polycrystalline films ($d \approx$ 470-1000 nm) prepared by chemical solution deposition (CSD – sol-gel technique) on (0001)-sapphire substrates by Fedorov *et al*. [24] in the temperature range 300-920 K. In the case of PTO, the temperature behavior of the *E*-symmetry component of the SM, as well as the TO2 frequency, agreed very well with those from Raman spectra on the bulk crystal. Also for both the PZT films, partial softening and hardening of the overdamped SM was revealed with (slightly diffuse) $T_C$ near 670 and 600 K for PZT-47 and PZT-25, respectively, in agreement with data on the bulk ceramics. In addition to SMs in the 1-3 THz range, relaxational CMs were revealed from the BWO spectra below ~0.5 THz in all the films, also needed to explain the microwave (MW) permittivity. Since the films are polycrystalline, the effective dielectric response is a mixture of the two principle dielectric functions along the optical *c*-axis and perpendicular to it. This mixture was successfully modeled for PTO at room temperature using the Bruggeman effective medium approximation (EMA) [25] with the correctly assigned phonons from the single-crystal Raman data [26].

An interesting phonon behavior in PZT (x = 0.6) nano-islands (~20 nm thick and ~100 nm in diameter) grown by CSD on MgO substrates was observed in the FTIR reflectance spectra [27]. Due to the IR depolarizing *E*-field on the nano-island boundaries, all phonons shift to higher frequencies, but the most drastic stiffening udergoes the SM, shifting from ~60 to 430 cm$^{-1}$. High-temperature reflectance studies revealed its only small gradual softening on heating towards the highest temperature of 870 K [27]. Recently, other CSD PZT-48 films on MgO and sapphire substrates were studied by BWO-FTIR spectroscopy at room temperature and compared with an amorphous film of the same composition [28]. Whereas the SM in the 20 cm$^{-1}$ range in the polycrystalline films contributed by 200-300 to the permittivity, the modes in the amorphous film were by an order of magnitude weaker.

The next attention was devoted to the incipient FE STO films, where the SM is known to be well underdamped in the bulk, including ceramics [29]. The first measurements were performed on the polycrystalline CSD film ($d$ = 275 nm) on sapphire substrate [30] with the surprising result that the SM softening ceases below ~100 K with its frequency remaining above 60 cm$^{-1}$, unlike in the bulk crystal where it softens down to ~10 cm$^{-1}$. Similar result was obtained also on several epitaxial films ($d \approx$ 500-2000 nm) prepared by pulsed laser deposition (PLD) on STO single crystal substrates with SrRuO$_3$ buffer layers ($d$ = ~350 nm) to screen the signal from the substrate, using the above mentioned FIR ellipsometric measurements on synchrotron [17], and was only qualitatively ascribed to possible defects and strains. More quantitative understanding of this effect was achieved later on by studying other two polycrystalline CSD STO films on (0001)-sapphire substrates [31] ($d$ = 360 and 720 nm), which showed qualitatively similar SM behavior (see Fig. 1a). However, the low-temperature SM frequency differed, it was ~60 and ~80 cm$^{-1}$ for the thinner and thicker film, respectively, even if they were prepared by the same technique and showed the same microstructure (columnar ~100 nm grains). However, in the thicker film small cracks along some of the grain boundaries were revealed. Applying the brick-wall (brick-layer) model, it was shown that a possible very small concentration of cracks, 0.2 and 0.4% for the thinner and thicker film, respectively, may account for the observed SM stiffening and permittivity reduction. Our later SM studies on fine-grain [32,33] and nano-grain fully-dense STO ceramics [34] have also revealed a similarly stiffened SM response, which was explained by very thin (~1 nm) dead layers of very low permittivity (~10) in the grain boundaries, presumably due to oxygen reduction, which, via the depolarizing field effects stiffen the SM and reduce the dielectric response. Moreover, the grain boundaries appear to be charged, which polarizes a thin layer along the grain boundaries (~3 nm) in the grain bulk and provides activation of the SM in the Raman response. Since the FIR and THz techniques probe the in-plane response (as well as predominantly the ellipsometry), the grain boundaries in films are expected to play a similar role as in the ceramics.

In Ref. [31] yet another STO film was studied, prepared by injection metalorganic chemical vapor deposition (MOCVD, $d$ = 290 nm), which appeared to be hetero-epitaxial, consisting of fully (111)-oriented ~100 nm grains rotated either by +30 or -30° with respect to the $c$-axis of the (0001) sapphire substrate [35]. Since the lattice misfit between the sapphire substrate and STO film is enormous (the inter-atomic spacing in the STO is by 13% larger than in the sapphire), the good film orientation is surprising and its small tensile strain (~0.2%) is determined only by the differences between the lattice expansion of the film and substrate (processing of the film was at ~800°C). It resulted in an in-plane FE PT near 125 K combined with the antiferrodistortive PT, revealed by the SM frequency minimum at $T_C$ and its splitting into three components below $T_C$ (see Fig. 1b). This appears to be the first reliable report on a FE PT in strained STO films, even if some reports on FE PT in STO films were published earlier without analyzing the strain [36].

The boom with FE PT in strained STO films was later initiated by Haeni *et al*. [37], who revealed an in-plane FE PT near 290 K (slightly diffuse) in a film grown by molecular–beam epitaxy (MBE) ($d$ = ~50 nm) on (110)-DyScO$_3$ substrates with an in-plane tensile strain of 0.8%, in agreement with theoretical calculations. Its dielectric response at 10 GHz was also well tunable by a bias $E$-field at room temperature. The strong tunability was then confirmed by THz spectroscopy on similar films grown by PLD [38,39] and it was shown to be caused by stiffening of the SM + CM response under the $E$-field. The FE PT was finally confirmed [40] on similar MBE STO/DyScO$_3$ films ($d$ = 100 and 30 nm) by the minimum in the SM frequency near $T_C \approx$ 280 K, accompanied by the appearance of CM close above $T_C$ which strongly contributed to the dielectric response on continuous softening even below $T_C$ down to ~150 K. Below $T_C$ two other SMs appeared in the FIR spectra with about the same softening temperature, assigned to the SM doublet from the R-point of the Brillouin zone due to the antiferrodistortive PT. The THz studies were continued by a thorough temperature and tunability study of epitaxial STO/DyScO$_3$ heterostructures, consisting of several alternating STO and DyScO$_3$ multilayers of different thicknesses [39,41,42], but most studies were performed on a heterostructure consisting of four couples of alternating layers with each layer thickness $d$ = 50 nm [43]. It was confirmed that the CM dominates the dielectric response below $T_C \approx 270$ K and the complete THz response down to 20 K can be fitted with a coupled SM oscillator – CM Debye relaxation model with hardening of the bare SM frequency and strengthening of the bare CM below $T_C$. All other parameters in the coupled oscillator-relaxation formula were left temperature independent [43]. The accurate behavior of the CM below $T_C$ and under bias field, however, would require also MW data. Study of a non-stoichiometric $Sr_{1-x}TiO_{3+\delta}$ film grown on DyScO$_3$ revealed broadening of distribution of the relaxation frequencies describing the CM, and a relaxor-like peak appeared at $T_C$ in $\varepsilon'(T)$ [44]. Ideal stoichiometric STO film revealed a single relaxation frequency, higher permittivity and larger SM softening near $T_C$ [44].

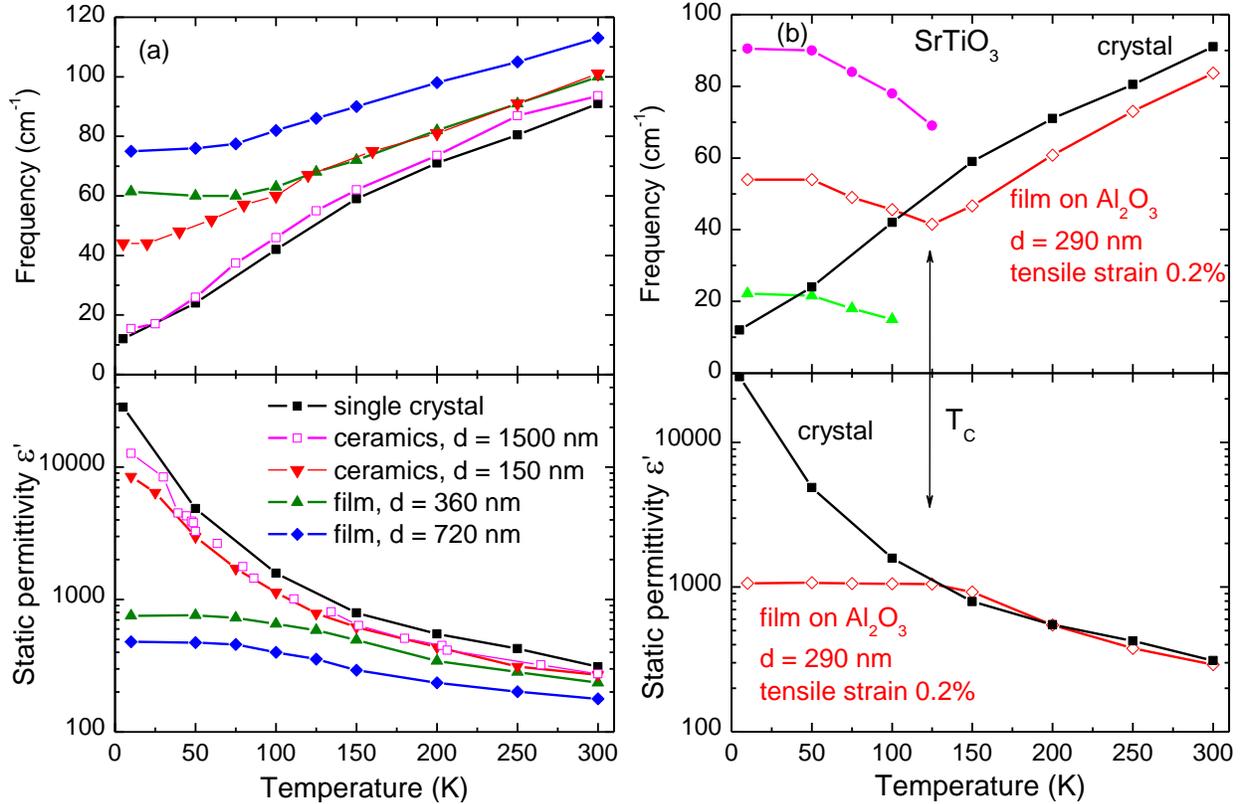

**Fig. 1**. Comparison of the temperature dependences of the SM frequency and static permittivity (obtained as the sum of phonon contributions) in STO single crystal with (a) STO ceramics of different grain size and unstrained polycrystalline CSD thin films. The grain boundaries and cracks stiffen the SM frequency and therefore the static permittivity is reduced. (b) Comparison of the SM and static permittivity behavior in STO single crystal with the hetero-epitaxial film (0.2 % tensile strained), which exhibits splitting of the SM below the FE PT. Modified from [31].

The spontaneous polarization in STO/DyScO$_3$ films lies in the film plane, but for possible applications in microelectronics the switchable out-of-plane polarization is required. This was predicted in compressively in-plane strained STO films [37], expected for epitaxial films on LSAT and NdGaO$_3$ substrates. Since both the FTIR and THz techniques probe the in-plane (non-FE) response, for such films presumably only SM stiffening can be expected and the static dielectric response is expected to be smaller without pronounced maxima at $T_C$. This was actually observed in films on both the substrates [37,45,46,47], without any anomaly in the SM behavior (see Fig. 2). In Ref. [45] a very small additional mode at 318 cm$^{-1}$ was revealed below ~150 K, assigned to IR activation of the silent $F_{2u}$ mode (its $E$-symmetry component) due to the FE PT. However, in Ref. [47] it was shown that, due to the probing unpolarized light, this was an artifact caused by the strong in-plane optical anisotropy of the NdGaO$_3$ substrate. Careful IR reflectance measurements in polarized light on (001) STO/(110) NdGaO$_3$ film ($d$ = 17 nm), performed also for the bare NdGaO$_3$ substrate, revealed even a splitting of the stiffened in-plane SM (Fig. 2), caused by the anisotropy of the in-plane film strain (mean compressive strain value of 1.1%), but no evidence for the out-of-plane FE PT was found, neither in the out-of-plane lattice parameters from the XRD data. This is in agreement with Ref. [48], where only relaxor FE behavior with a broad out-of-plane permittivity maximum at 100-150 K was observed in a similarly compressed film. On the other hand, very recently a FE PT near 140 K was observed in MBE (001) STO/LSAT ($d$ = 160 nm) film compressively strained by 0.95% [49], but no SM studies for the out-of-plane orientation are available. Very recently the SM in the same type of STO/LSAT films was studied by the FIR ellipsometry including very small thicknesses ($d$ = 82 and 8.5 nm) [50] with the very similar stiffened SM temperature dependence as in Ref. [46] ($d$ = 49 and 107 nm), indicating that the ellipsometry probes also only the in-plane polarized phonons.

In Fig. 2 we have summarized our FTIR data on SM and CM in strained epitaxial STO films grown on various substrates compared with the SM in single crystal. The differences are really impressive and reach more than one order of magnitude in the SM frequencies. Let us note that the variations in the higher-frequency TO phonons are much smaller and do not show temperature anomalies near $T_C$ [40,46,47]. Nevertheless, frequencies of these hard phonons are also slightly tunable by electric field, which has been seen in IR reflectivity spectra of STO/DyScO$_3$ [51]. Finally, we would like to note that the ferroelectricity in STO can be induced even in strain-less homo-epitaxial films (grown on (001)-STO substrates) due to a non-stoichiometry, mainly Sr-deficiency [52].

STO is the $n = \infty$ member of the Sr$_{n+1}$Ti$_n$O$_{3n+1}$ family. Bulk Sr$_{n+1}$Ti$_n$O$_{3n+1}$ crystallizing in Ruddlesden-Popper structure, exhibits incipient FE behavior and its permittivity increases with the number of perovskite layers $n$ [53]. In strained Sr$_{n+1}$Ti$_n$O$_{3n+1}$ films grown on DyScO$_3$, FE PTs were observed in samples with $n \geq 3$ [54] with $T_C$ increasing with rising number of perovskite layers $n$. THz transmission and IR reflectivity spectra reveal the FE SM, which drives the FE PT, but near $T_C$ the dielectric strength of CM dominates [55]. The $n = 6$ film (Sr$_7$Ti$_6$O$_{19}$) shows $T_C$ = 180 K, high $E$-field tunability of permittivity and exceptionally low dielectric loss at 300 K that rivals all known tunable MW dielectrics [54]. Analysis of the MW, IR and THz spectra shows that the low room-temperature MW loss in this film has the origin in the absence of CM at 300 K, while the high permittivity tunability is due to sensitivity of the SM to the $E$-field [55].

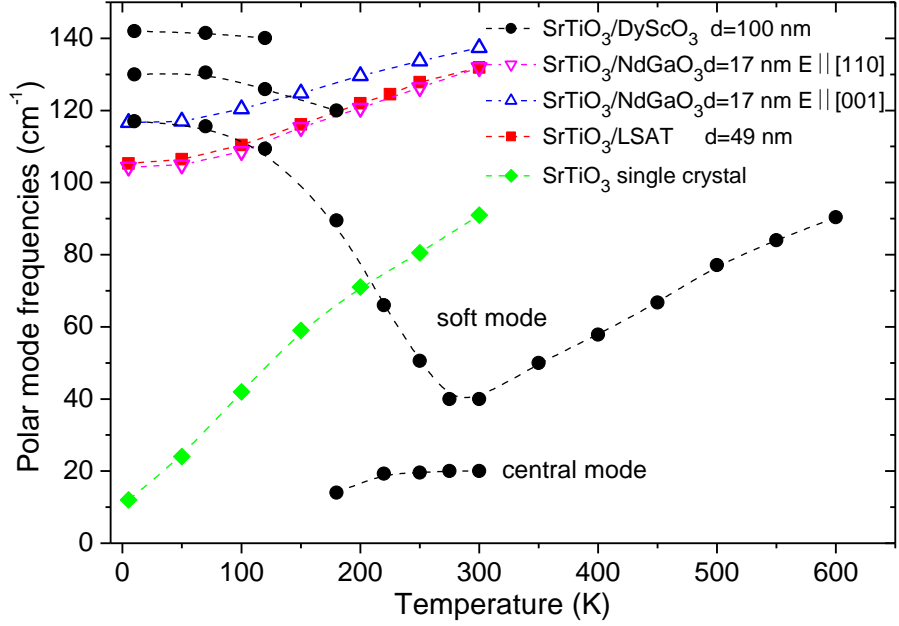

**Fig. 2.** Temperature dependences of the SM frequencies in strained STO films grown on various substrates (modified from [11]). Compressively strained films exhibit stiffening of the SM with respect to the crystal, while tensile strained STO/DyScO$_3$ exhibits SM anomaly and appearance of the CM near 280 K due to a strain-induced FE PT. Below ~180 K two new modes activate above 120 cm$^{-1}$ in this film due to an antiferrodistortive PT within the FE phase.

The second most popular incipient FE is KTaO$_3$ (KTO), where only two publications were devoted to the IR and THz SM studies [56,57]. In Ref. [57] the CSD polycrystalline (grain size ~160 nm) KTO film ($d \approx 200$ nm) on (0001) sapphire substrate, partially (001) oriented, was shown to be FE with diffuse $T_C \approx 60$ K. The reason was shown to be the compressive strain due to larger thermal expansion of the sapphire compared to KTO (the film was annealed at 900°C), which could induce a FE phase with preferential out-of-plane orientation of $P_s$. In spite of it a minimum in frequency of the in-plane polarized SM was observed near $T_C$ [57], while in crystal and ceramics the SM gradually softens on cooling and levels off at ~20 cm$^{-1}$ below 20 K [58,59].

There are more materials showing up the incipient FE behavior (we shall discuss some of them below among multiferroics), but their SM behavior was rarely studied in thin films. The first FTIR SM study (earlier than our first FE SM study [24]) was performed on a series of semiconducting PbTe films of different free-carriers concentration ($5 \times 10^{16}$ - $5 \times 10^{17}$/cm$^3$) and various thicknesses ($d$ = 1-14 μm) grown on NaCl and BaF$_2$ substrates [60]. SM softening was detected from ~32 cm$^{-1}$ at 300 K to 18-22 cm$^{-1}$ at 5 K (only slightly dependent on the free-carrier concentration), but the dielectric strengths were not evaluated. More recently, the related semiconducting Pb$_{1-x}$Eu$_x$Te epitaxial thin film system ($0 \leq x \leq 0.37$) was studied in the whole IR-THz range on BaF$_2$ and Si substrates from 300 down to 5 K [61]. The SM behavior was very similar to that in the PbTe films and their dielectric strengths varied from ~250 at 300 K to ~1000 at 5 K.

The most popular and studied FE is barium titanate BaTiO$_3$ (BTO) with the FE $T_C \approx 400$ K. The SM response in crystals is rather broad and consists of two overlapping overdamped modes in the paraelectric phase [62], which split into well resolved CM and rather high-frequency $A_1$ symmetry SM in the response along the tetragonal FE axis below $T_C$ [63]. The SM response in the low-temperature FE phases (orthorhombic and rhombohedral) in crystals is not accurately known, since no single-domain samples in these phases were studied. The effective (averaged over the anisotropy in the FE phases) SM+CM frequencies (denoted by TO1) and of all other polar phonons in BTO coarse-grain ceramics in all the phases are shown in Fig. 3a [64,65]. In Fig. 3b the temperature dependence of the total polar phonon contribution to the low-frequency permittivity is plotted for BTO ceramics with various grain size (10000, 1200, 100 and 50 nm). The pronounced grain-size effect on the permittivity values is due to a low-permittivity grain-boundary layer as discussed for the STO ceramics [64] and is caused mainly by stiffening of the SM. Note that in the paraelectric phase the phonon response yields the whole directly measured permittivity values, whereas in the FE phases other lower-frequency dispersion contributes to the dielectric response (domain-wall displacements, piezo-effect on grain boundaries).

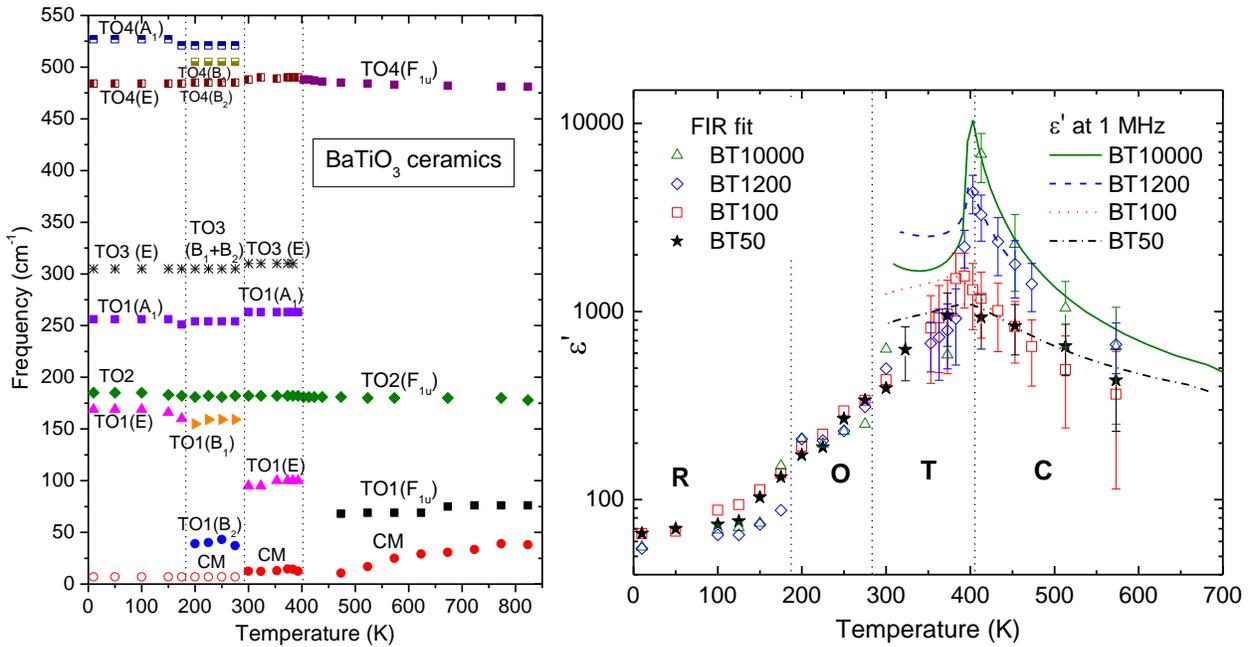

Fig. 3. (a) Temperature dependence and assignment of the TO polar mode frequencies in BTO coarse-grain ceramics. Modified after [65]. (b) Temperature dependence of the low-frequency permittivity of BTO ceramics with various mean grain sizes (indicated in the figure in nm), calculated form the phonon response, compared with directly measured permittivity at 1 MHz. After [64].

The first SM studies of BTO thin films were performed on (0001)-sapphire substrate, grown by MOCVD ($d$ = 200 nm) and CSD ($d$ = 375 and 750 nm,) [66,67,68,69,70]. In all the cases, the TO1 phonon did not reveal any significant anomaly at the expected $T_C$ [71,69] – see Fig. 4. It only splits below $T_C$ and its $A_1$ component jump-wise appears above 220 cm$^{-1}$ in a similar way as in the BTO ceramics and crystals (compare Fig. 3a and 4a). CM markedly hardens below $T_C$ and therefore also the static permittivity decreases on cooling. Note the one order of magnitude lower paraelectric permittivity in the BTO film than in the ceramics. Similar to nanoceramics, this is caused by stiffening of the SM and lower dielectric strength of the CM in the films. Raman spectra can generally distinguish between paraelectric and FE phase in perovskites, since the

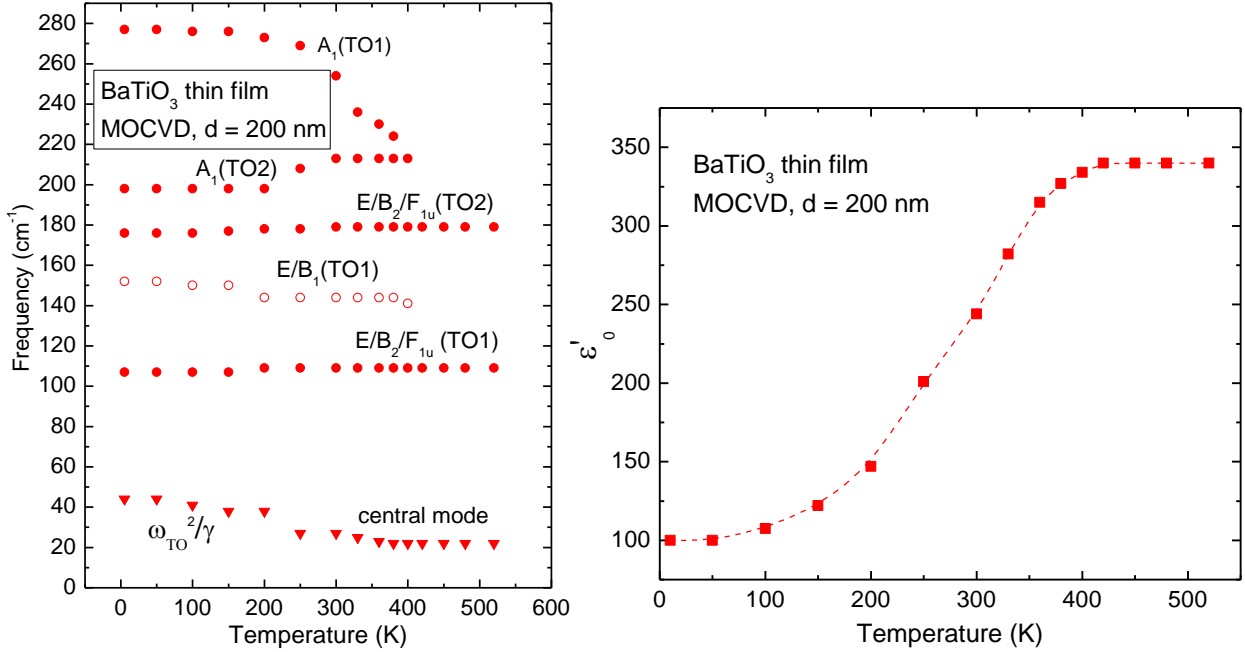

Fig. 4 (a) Temperature dependence of the TO mode frequencies in BTO hetero-epitaxial film ($d$ = 200 nm). CM is overdamped, therefore frequency of dielectric loss peak $\varepsilon''(\omega)$ corresponding to $\omega_{TO}^2/\gamma$ is plotted. (b) Temperature dependence of the static permittivity calculated from the sum of phonon contributions. The observed decrease in permittivity is mainly caused by the CM hardening on cooling and its decrease in the dielectric strength. Modified after [71].

silent TO3 mode (at ~312 cm$^{-1}$ in BTO) is inactive in the paraelectric phase and activates in the FE phase [6]. Our Raman spectra of polycrystalline BTO films shown that the FE phase remains stable at least up to 550 K ( the highest measured temperature) [69].

The most thorough spectroscopic studies of BTO thin films (but without the low-frequency SM+CM spectral range) were carried out using the ultraviolet Raman scattering by Tenne *et al.* (see [72] and the review in [12]). First they studied thicker films on STO and LaAlO$_3$ substrates with SrRuO$_3$ buffer layers ($d \approx 1000$ nm), which show a small in-plane compressive strain due to a smaller thermal expansion of the substrates. This induced the in-plane orthorhombic FE phase down to low temperatures, in agreement with the theory. In a more recent paper Tenne *et al.* [73] studied very thin (001) epitaxial films on STO substrates. The 10-nm BTO film was fully in-plane compressively strained (2.2%) and showed the out-of-plane *c*-axis ferroelectricity up to $T_C \approx 925$ K. In thinner films, $T_C$ continuously decreased and reached ~70 K in the thinnest 1.6-nm film due to the size effect of the out-of-plane ferroelectricity, in agreement with the theory [73]. It appears that the films show the tetragonal FE phase down to low temperatures, the other two FE PTs, known from single crystals, being absent.

BTO-STO (BST) solid solution is the best studied mixed crystal system, where the SM+CM spectra of bulk ceramics were recently published for the whole composition range (see [74] and references therein). In good ceramics the FE PTs remain sharp in almost the whole composition range (above ~10% Ba) and the CM gradually strengthen in the THz spectra on increasing the Ba content [74,75,76,77]. The first thin-film spectra were published for 10% of Ba [67,68] on similar type of MOCVD and CSD films as discussed for STO and BTO. As for BTO, the SM did not soften as much as in the bulk ceramics and $T_C$ was shifted from ~80 up to ~150 K, presumably due to a strain. In Ref. [78], SM+CM spectra for CSD BST films with 50, 70 and 80% of Ba ($d$ =

500, 1000 and 300 nm, respectively) were studied using the THz +FTIR transmission for 100-420 K. Two overdamped CMs were revealed in the 10 and 30 cm$^{-1}$ range with no appreciable softening near $T_C$ and much weaker dielectric strengths compared to the bulk. The PT temperatures $T_C$ were revealed only by weak maxima in the CM strengths. The two CMs were assigned to CM shifts (in part of the sample) from the bulk values due to the depolarizing field effects on the grain boundaries and other inhomogeneities in the microstructures of the films. More detailed studies were performed using Raman scattering on PLD films ($d \approx 300$ nm) on STO substrates with SrRuO$_3$ buffer layers for Ba composition up to 50% [12]. All the films showed smearing of the PTs with broad overdamping of the SM and indicated relaxor-type behavior due to the relaxation of the selection rules for the forbidden modes, in analogy with similar STO and BTO films. The probable reasons for this behavior were suggested to be the charged grain boundaries as in the STO ceramics and nano-ceramics [29,32,33,34], and the strain inhomogeneities.

In Ref. [79] the thickness dependence (from 10 to 420 nm) of the room temperature SM parameters were studied by BWO-FTIR spectroscopy in rf sputtered BST films with 80 and 85% Ba on MgO substrates. Whereas for the films above 30 nm no pronounced differences were found in the very broad SM+CM response with the dielectric strengths of 300-400, for the 10-nm film a pronounced strengthening of the SM was revealed up to $\Delta\varepsilon \approx 1400$, assigned to tensile in-plane stress, which presumably lowers the $T_C$ close to room temperature. In Ref. [80] similar films of BST with 70% Ba ($d = 36$ and 800 nm) on MgO substrates were studied using the same technique. The broad SM+CM response was described by 2 oscillators in the 70 and 40 cm$^{-1}$ range with stronger dielectric strength (~570) of the thinner film. It was emphasized that for a proper evaluation of the SM strength a careful fit of the bare substrate (including weak multiphonon features) is essential. In a more recent paper [81], broader thickness dependence ($d = 36$-1500 nm) in the SBT films with 80% of Ba on MgO substrates was studied by the BWO-FTIR spectroscopy with the film fitting both by classical as well as generalized oscillators. Dielectric strengths of the SM+CM were shown to behave non-monotonically, on increasing thickness above 60 nm they first decreased up to 800 nm and then increased.

The last type of films based on BTO and STO are BTO/STO superlattices. The effective SM behavior for the (BTO)$_8$/(STO)$_4$ superlattices repeated 40 times (total thickness $d = 192$ nm) on DyScO$_3$, SmScO$_3$, TbScO$_3$ and EuScO$_3$ substrates was studied by Železný et al. [82,83,84] using the polarized FTIR reflectance. In case of the film on SmScO$_3$ substrate also the temperature dependence was studied [83], but no FE PT was detected from the phonon behavior. The effective SM was split due to the substrate anisotropy and hardened gradually on cooling, up to rather high frequency ~150 cm$^{-1}$ at 10 K. Also the Raman spectroscopy can be used as a sensitive tool for the FE order, because the polar phonons in crystals with centrosymmetric paraelectric phase become Raman active only in the FE phase. Unfortunately, substrate phonons usually dominate in the Raman spectra of thin films, because penetration depth of the laser beam is typically several micrometers (depending on the band structure of the thin film). Nevertheless, the penetration depth is strongly reduced for the UV radiation. For that reason Tenne et al. [72] used UV Raman spectroscopy for the study of various BTO/STO superlattices grown on the STO substrate. They found that the BTO layers are FE even when their thickness is only one unit cell (0.4 nm) and that they can induce polarization in the adjacent paraelectric STO layers. $T_C$ was tuned from ~150 up to ~640 K.

**Relaxor ferroelectric perovskites**

We assume here that the reader is familiar with the concept of relaxor FEs (briefly called relaxors). It is well known that in bulk relaxor FEs the main dielectric dispersion of the giant dielectric response occurs below the phonon range, i.e. in MW and lower frequencies, strongly dependent on the temperature [5,6,7,8]. It can be usually described by a relaxation, which appears below the Burns temperature $T_d$ (temperature below which the polar nanoregions (PNRs) are supposed to appear), broadens on cooling and its characteristic frequency follows the Vogel-Fulcher law. This causes the well-known frequency dependent maximum of the dielectric response above the Vogel-Fulcher temperature $T_{VF}$. The SM in relaxors is known to soften only partially and contributing only by a small amount (of the order of several hundreds) to the total dielectric permittivity, which is typically several tens of thousands in its maximum.

The first SM studies in thin film relaxors were performed on PZT doped with La [$(Pb_{1-x}La_x)(Zr_yTi_{1-y})O_3$ - PLZT 100($x/y/1-y/$)] of the classical relaxor composition PLZT 9.5/65/35 [85]. Similarly to PZT [24], the polycrystalline CSD films ($d$ = 139 nm) were prepared on (0001) sapphire substrates and the transmission FTIR spectra were combined with those from BWO to evaluate the complete SM+TO2 response from 4 to 350 cm$^{-1}$ range and compare it with the data from FTIR reflectivity on bulk ceramics. Reasonable agreement of the SM parameters (only slightly higher frequency and damping) with those of bulk ceramics was obtained. As in the bulk, the SM undergoes a small gradual softening (together with an overdamping) on heating from 8 to 523 K and shows no anomaly near the maximum of the dielectric response (~400 K). The BWO data indicate also a CM-type relaxation below the SM, but the accurate evaluation requires additional MW and lower-frequency data. To account for the low-frequency data at room temperature, the relaxation frequency of ~6 GHz was needed and its dielectric strength was more than by an order of magnitude higher than that of the SM. Later on, also other PLZT film compositions (CSD films of PLZT 8/65/35 and 2/95/5 on sapphire substrates) were studied by the FTIR transmission [86,87] and similar results for the SM behavior were obtained. Moreover, from the phonon spectra it was possible to detect the presence of the pyrochlore secondary phase, if present in the films.

The best investigated relaxor is $PbMg_{1/3}Nb_{2/3}O_3$ (PMN). The SM+CM response on a CSD film ($d \approx$ 500 nm, grain size ~60 nm, prevailing (111) orientation) was first time studied by the FTIR transmission in Ref. [88] in a broad temperature range 20-900 K and compared with the crystal data. In Ref. [89] also lower-frequency data for films, crystals and ceramics were summarized. Underdamped SM was present in the whole temperature range with a partial softening on heating reaching minimum frequency of ~50 cm$^{-1}$ in the 600 K range and above. The thin film response of the SM agrees quite well with that of the single crystal, but differences are observed for the lower-frequency CM dispersion, the thin film showing much smaller permittivity and therefore smaller CM strength. This can be understood as due to a partial clamping of the PNR response at the grain boundaries, as in nanoceramics [90]. Recently, the SM response in PMN single crystals was studied by hyper-Raman scattering [91] and an additional component of the SM was detected at lower frequencies (20-35 cm$^{-1}$), interpreted as the *E*-component of the split SM (perpendicular to the local dipole moment of the PNR) due to strong dielectric anisotropy of the PNRs, suggested in Ref. [92]. This excitation is also seen in the THz range and remains present even at low temperatures, where the PNRs are frozen. Its frequency corresponds to the overdamped mode originally assigned to CM [88] – see Fig. 5a. In Fig. 5b we show that below $T_d \approx$ 620 K a broad relaxation develops in PMN single crystals below the *E*-component of the SM and broadens on cooling. The low- and high-frequency edges of the relaxation time distribution can be assigned to flipping and breathing frequencies of the PNRs. Temperature dependences of both the edges were determined from the broad-band dielectric spectra of ceramics [88, 93]. The flipping mode follows the Vogel-Fulcher law and freezes below the freezing temperature $T_f$, while the breathing mode follows the Arrhenius behavior.

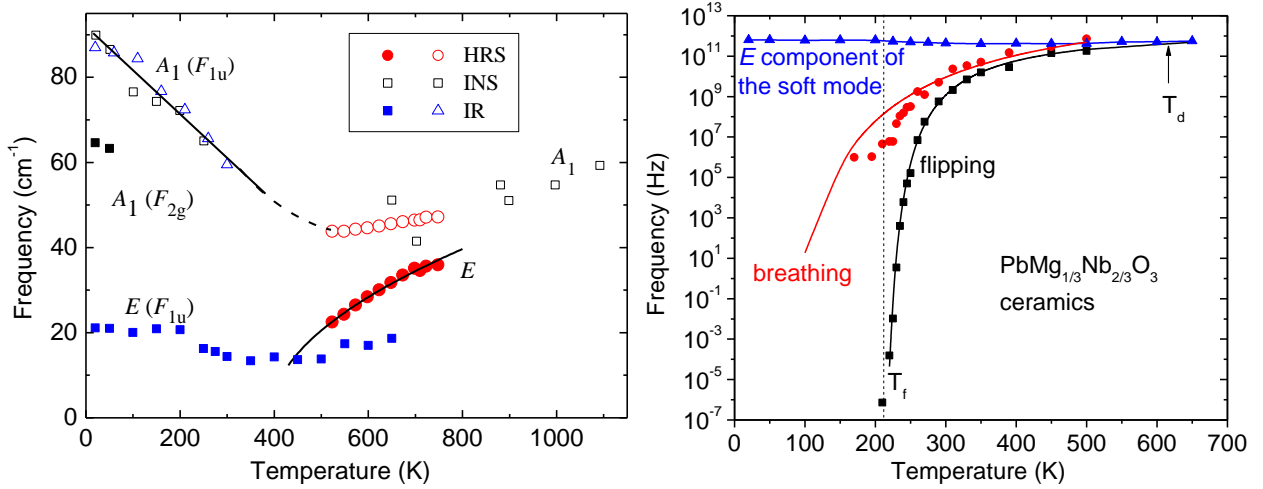

**Fig. 5.** (a) Temperature dependence of $A_1$ and $E$ component of the $F_{1u}$ SM in the cubic phase of PMN, determined from the IR spectra of the thin film [88] and hyper-Raman scattering (HRS) [91] in the crystal. The same SM frequencies are seen. (b) Below $T_d \approx 620$ K a relaxational CM appears. The distribution of its relaxation frequencies broadens on cooling – see text. Modified after [93].

PMN is intrinsically chemically disordered due to its B'-B''-site stoichiometry of 1/3 and 2/3 and no correlation was found between the PNRs and possible chemical nano-order. This differs from the complex perovskites of ½ and ½ B-site stoichiometry, which can be, in some cases, prepared in variously ordered forms [94]. The best example is $PbSc_{1/2}Ta_{1/2}O_3$ (PST), which shows relaxor behavior in the disordered form, but undergoes a rather sharp FE PT ($T_C \approx 300$ K) in the ordered form [95]. IR reflectivity of PST ceramics was shown to be a very sensitive technique to determine the degree of the B-site order due to the strength of the B'-B'' stretching mode near 315 cm$^{-1}$ [94] and the corresponding detailed FTIR studies were performed in Ref. [96]. Due to the absence of SM softening (~60 cm$^{-1}$) and presence of the relaxational CM at ~12 and ~30 cm$^{-1}$ above and below $T_C$, respectively, it was concluded that the FE PT is of the order-disorder type, connected with ordering of the off-centered Pb ions. This was supported also by MW measurements, which revealed another relaxation below $T_C$ in the ~0.5 GHz range [97]. Also in the thin film form it is possible to prepare variously ordered samples and the IR transmission spectra of PST films ($d = 500$ nm) on (0001) sapphire substrates prepared by CSD with various annealing up to the 78% order were studied in Ref. [98]. Qualitatively similar results were obtained for the SM behavior with no pronounced differences between the ordered and disordered samples. Nevertheless, at 100 kHz, the ordered PST ceramics exhibit the same permittivity as the ordered thin films, while the disordered PST ceramics show much higher permittivity than the thin films [99]. It means that in disordered samples the dielectric strength of the CM in ceramics is much higher than in the thin films. The SM frequency stays near 45 cm$^{-1}$ above ~600 K and hardens below it, particularly below $T_C$. The intrinsic CM, which appears below ~600 K and softens towards $T_C$, was seen only above $T_C$ and was assigned to dynamics of the PNRs. Existence of the mode near 30 cm$^{-1}$ below $T_C$ down to the lowest temperatures was assigned to the *E*-component of the split SM like in the case of PMN.

B-site order can be obtained also in PbMg$_{1/3}$Ta$_{2/3}$O$_3$ (PMT) by a proper high-temperature heat treatment or by a small admixture of PbZrO$_3$. 15% and 90% ordered PMT ceramics and disordered PMT films grown on sapphire were investigated from 100 Hz up to 100 THz [100]. Similar behavior of the CM as in PMN (Fig. 5) was observed: it appears below $T_d$ near the SM frequency and highest and lowest relaxation frequencies (expressing flipping and breathing modes) follow the Arrhenius and Vogel-Fulcher laws, respectively. The degree of the B-site order in the PMT ceramics has only a small influence on the parameters of the dielectric relaxation and almost no influence on the phonon parameters. Also the phonons in ceramics and thin films are equal to each other within accuracy of the measurements. SM is split in two components at all investigated temperatures (10 – 900 K) indicating a lower local symmetry than the cubic one. Since the SM is split also above $T_d$ were no PNRs are assumed, it indicates the role of random electric fields from the two B-site cations of different valences, which locally break the cubic symmetry [101].

**Other ferroelectrics**

Except for perovskite FEs, incipient FEs and relaxors, only few proper FE thin films with other structures were studied by IR-THz spectroscopy concerning their SM behavior. The most simple ferroelectric structure shows GeTe with a simple cubic rock-salt structure (β phase) and a proper (equi-translational) FE PT into a rhombohedral phase near 620 K (α phase). The material is also attractive as a so-called phase-change material which can be transferred into an amorphous glassy phase, but from the point of view of ferroelectricity it has a drawback of rather high semimetallic conductivity. In Ref. [102] a magnetron sputtered thin film ($d$ = 500 nm) on silica-glass substrates was studied by THz spectroscopy from room temperature up to ~725 K. As expected, the THz conductivity of 1400-2200 S/cm (in agreement with DC conductivity) dominated the spectra, but assuming the SM parameters below $T_C$ (in the 50-90 cm$^{-1}$ range) known from Raman scattering, it was possible to suggest the SM+CM behavior also above $T_C$. It was shown that the CM dominates in the dielectric spectra so that the PT appears to be predominantly of the order-disorder nature.

Another case, in which the thin film helped to study the SM+CM behavior, was bismuth titanate SrBi$_2$Ta$_2$O$_9$ (SBT), the famous material with Aurivillius structure used in FE memories due to its easy and fatigue-free polarization switching. The bulk SM properties were studied by FTIR reflectivity [103] and it was revealed that at the FE PT (~600 K) no anomaly in SM behavior was observed, only its gradual softening, and the SM strength could not explain the low-frequency permittivity. More precise THz and FTIR transmission measurements were performed on a rather thick PLD films ($d$ = 5500 nm) on (0001)-sapphire substrates [104,105,106] up to 950 K and in addition to SM also a relaxational CM was revealed below ~15 cm$^{-1}$ with softening near $T_C$ – see Fig. 6. Its dielectric strength was comparable to that of the SM so that the FE PT looks as more of order-disorder type, but the new first principles and Monte Carlo simulations revealed trilinear coupling of one hard zone-center phonon ($\Gamma_5^-$ symmetry) and two soft zone-boundary phonons ($X_3^-$ and $X_2^+$ symmetries), which drive the FE PT [107]. Above this hybrid triggered FE PT the SM is not IR active, which explains its absence in the IR spectra.

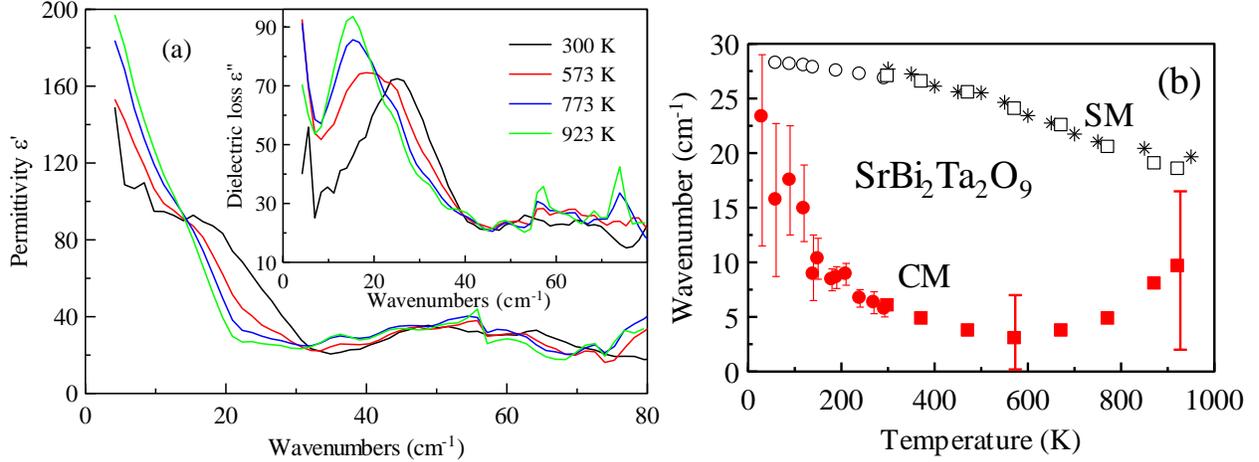

**Fig. 6**. (a) Complex dielectric spectra of the SBT thick film obtained from the THz spectroscopy. (b) Temperature dependence of the SM and relaxation CM frequency. Only the CM frequency exhibits minimum near $T_C \cong 600$ K, while the lowest frequency phonon gradually softens on heating without any anomaly at $T_C$. After [105].

Recently there appeared a report of switchable FE polarization in the 1.6 % tensile-strained $TiO_2$ thin film with anatase structure (a-$TiO_2$) using piezoelectric force microscopy (PFM) [108]. The measured polarization disappeared above 450 K, which was assigned to the Curie temperature $T_C$. However, Skiadopoulou *et al.* [109] performed XRD, second-harmonic generation (SHG) and FTIR investigations with the same film and none of the techniques revealed any hint of FE PT in this film. It was concluded that previously reported polarization is defect and/or electrochemically induced, in spite of the fact that it was stable in time at elevated temperatures. These results could be generalized to other dielectric systems: Observation of poling and FE hysteresis loops cannot be taken as conclusive evidence of the presence of ferroelectricity. Other experiments like XRD, SHG and lattice dynamic studies are necessary.

**Magnetoelectric materials**

An intensive effort is concentrated on the study of magnetoelectric multiferroics due to their wide range of future promising applications, such as information storage, sensing, actuation and spintronics. Unfortunately, most of them exhibit antiferromagnetic order and therefore only weak magnetoelectric coupling. Moreover, great majority exhibits multiferroic properties at cryogenic temperatures, which does not allow using these materials in technical applications. For that reason there is an intensive search for new multiferroics with high magnetoelectric coupling and critical temperatures above room temperature.

Fennie and Rabe [110] suggested theoretically a new route for preparation of multiferroics with a strong magnetoelectric coupling. They proposed to use a biaxial strain in epitaxial thin films for inducing the FE and ferromagnetic states in materials, which are in the bulk form paraelectric and antiferromagnetic. The basic condition for such a material is the strong spin-phonon coupling. Based on their first-principles calculations, Fennie and Rabe [110] proposed to use $EuTiO_3$, because this materials in its bulk form is incipient FE and its permittivity strongly changes at the antiferromagnetic PT due to the strong spin-phonon coupling [111]. Temperature dependence of the permittivity in $EuTiO_3$ crystal was successfully explained by a strongly anharmonic behavior of the SM [112,113]. The bulk $EuTiO_3$ ceramics undergoes an antiferrodistortive PT from the high-temperature cubic *Pm-3m* structure to the tetragonal *I4/mcm* phase near 280 K [114,115]. For that reason the SM splits in the tetragonal phase and both components of the SM soften on cooling. Lee *et al*. [116] grew the epitaxial MBE $EuTiO_3$ thin films on $DyScO_3$ substrates and due to the 1% tensile strain the SM (seen in the IR reflectance) is much softer than in the bulk, reaching a minimum frequency at $T_C$ = 250 K (Fig. 7a). Due to the SM anomaly the permittivity exhibits maximum (Fig. 7b), typical for displacive FE PTs. FE polarization was calculated from the first principles [116] and also estimated based on the value of $T_C$ [117], and the resulting value was 20 – 30 $\mu C/cm^2$. The non-centrosymmetric structure of the thin film was also proven by observation of the SHG signal below $T_C$. Magneto-optic Kerr effect studies revealed strain-induced ferromagnetic state below 4.3 K. All these studies confirmed theoretical predictions that the strain can induce FE and ferromagnetic order in paraelectric antiferromagnets. Nevertheless, the magnetic Curie temperature $T_C$ = 4.3 K is too low for practical applications, despite the FE critical temperature is close to room temperature and can be even enhanced by a higher strain. Higher magnetic and FE critical temperatures where theoretically predicted in the strained EuO [118], $CaMnO_3$ [119] and $SrMnO_3$ [120] thin films, but only in the tensile-strained $SrMnO_3$ grown on LSAT substrate this was confirmed from SHG [121]. $CaMnO_3$ grown with 2.3% tensile strain on $LaAlO_3$ exhibits only incipient ferroelectric behavior below 25 K [122]. Goian *et al*. [123] investigated several strained $SrMnO_3$ thin films by the FTIR reflection, but did not obtain reliable phonon parameters, presumably due to the high phonon damping in this material, observed also in $Sr_{1-x}Ba_xMnO_3$ ceramics.

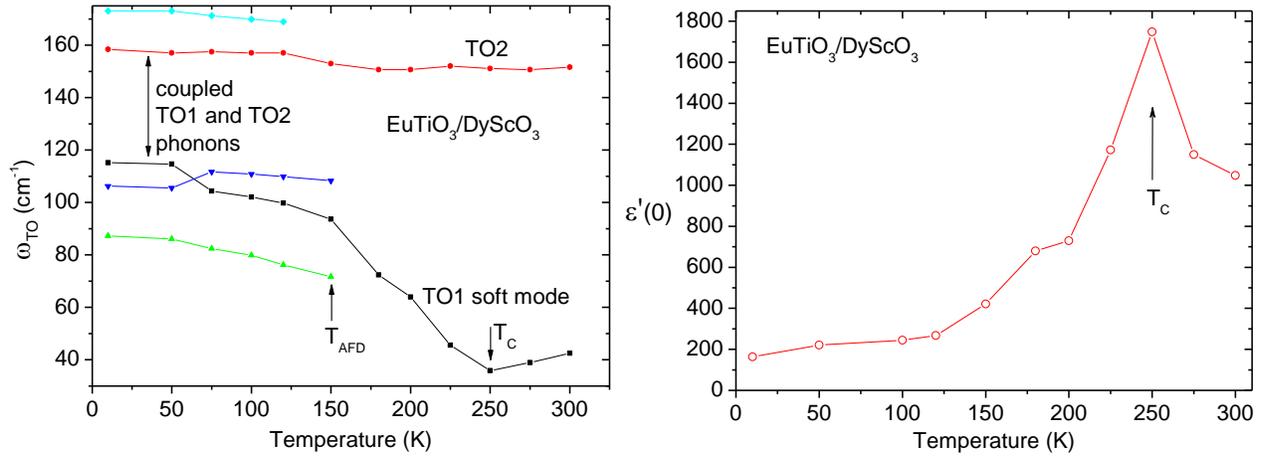

Fig. 7 (a) Temperature dependence of the lowest-frequency phonons and (b) static permittivity in the $EuTiO_3/DyScO_3$ film. SM anomaly induces dielectric anomaly at $T_C$ and two new modes activate in the IR spectra below 150 K due to the antiferrodistortive PT at $T_{AFD}$. Modified after [116].

Permittivity dramatically changes with magnetic field in $EuTiO_3$ [111]. Since the high value of permittivity is caused by the SM contribution, it is natural to expect that the SM frequency will change in the magnetic field. First attempt to see the SM tuning in magnetic field in ceramics [112] was not successful due to too broad reflectivity band. Nevertheless, in compressively strained $EuTiO_3$ films grown on LSAT substrates, the SM frequency is significantly stiffened above 100 $cm^{-1}$ [124] and shows a rather narrow reflectance peak, whose change with magnetic field and temperature can be studied with much higher accuracy than in the bulk samples.

Actually, reliable tuning of the SM frequency by 2 cm$^{-1}$ was detected in the FTIR reflectance spectra [124], explaining in this way the magnetodielectric effect in EuTiO$_3$.

One of the rare multiferroics with both magnetic and FE critical temperatures above room temperature is BiFeO$_3$. Comparison of polar phonon frequencies in bulk ceramics and thin film grown on (110) TbScO$_3$ was recently published by Skiadopoulou *et al*. [125]. Since the 300-nm film was strain-free, their phonon frequencies were almost the same as in the bulk samples, but the phonon damping was significantly lower in the epitaxial films than in the bulk ceramics. The spin excitations, seen in the bulk samples, were both IR and Raman active [126] so that they were assigned to electromagnons [125]. This fact was also confirmed by observation of directional dichroism in the FTIR transmission spectra [127]. Nevertheless, the spin excitations are very weak and were not detected in IR-THz spectra of the thin films [125]. The three lowest frequency phonons gradually soften on heating towards the FE PT at 1020 K, causing only a small increase in the static permittivity with temperature, in agreement with the improper character of the FE PT (note that the unit cell doubles at the paraelectric to FE PT). Related BiFeO$_3$ rf-sputtered films on MgO substrates doped with 2% of Nd of different thicknesses (32-500 nm) were studied at room temperature using BWO-FTIR techniques [128] as in Refs. [13,79,80,81]. The heavily damped SM was revealed in the 30 cm$^{-1}$ range (not seen in thick films and bulk ceramics) with the dielectric strength increasing with decreasing thickness up to ~400 for the 32-nm film. The effect, not observed in pure BiFeO$_3$ films [125], was assigned to structural distortions in thinner Nd-doped films.

Also the B-site ordered double-perovskite Bi$_2$FeCrO$_6$ exhibits multiferroic properties above room temperature [129]. A thin film grown on LaAlO$_3$ (*d* = 600 nm) was investigated using the FTIR reflectance spectroscopy from 10 to 900 K. [129] Gradual softening of several phonon frequencies with rising temperature was observed, which explains the gradual increase in permittivity from 50 (at 10 K) to 110 (at 900 K), but no signature of FE PT was observed, lying probably above 900 K. Nevertheless, around Néel temperature of ~600 K more remarkable phonon softening was seen, probably due to a spin-phonon coupling.

**Conclusions**

The FIR and THz SM and CM spectroscopies are shown to be very sensitive techniques to study the displacive FE and related thin films on dielectric substrates from several µm down to ~10 nm film thicknesses. Unlike Raman spectroscopy, it can be usually used also in the paraelectric phase and the evaluated dielectric strengths of the SM+CM enable to estimate the temperature dependences of the low-frequency dielectric properties in the film plane including the possible existence of FE PTs without using electrodes. Comparison with bulk materials shows that the SM properties in thin films are very sensitive to the in-plane strain in epitaxial films and to grain boundaries in polycrystalline films mainly due to possible depolarizing *E*-field effect caused by the probing wave at grain boundaries on the SM+CM response. The SM damping is also very sensitive to the film homogeneity of the strain. It appears that the discussed techniques are quite efficient and should be used more frequently for characterization of the FE thin film properties.

**Acknowledgement**

This work was supported by the Czech Science Foundation, Project 15 – 08389S. The authors thank T. Ostapchuk, V. Bovtun, F. Kadlec and D. Nuzhnyy for the help with figure preparation.